# The Spatiotemporal Evolution of Temperature During Transient Heating of Nanoparticle Arrays

**Chen Xie**[1]

Department of Mechanical Engineering, University of Texas at Dallas

800 West Campbell Road, Richardson, Texas 75080, United States.

Chen.Xie@utdallas.edu

**Zhenpeng Qin***

Department of Mechanical Engineering, Department of Bioengineering, Department of Bioengineering, Center for Advanced Pain Studies, University of Texas at Dallas; Department of Surgery, University of Texas at Southwestern Medical Center.

800 West Campbell Road, Richardson, Texas 75080, United States.

Zhenpeng.Qin@UTDallas.edu

* Corresponding author

## ABSTRACT

*Nanoparticle (NP) are promising agents to absorb external energy excitation and generate heat. Cluster of NPs or NP array heating have found essential roles for biomedical applications, diagnostic techniques and chemical catalysis. Various studies have shed light on the heat transfer of nanostructures and greatly advanced our understanding of NP array heating. However, there is a lack of analytical tools and dimensionless parameters to describe the transient heating of NP arrays. Here we demonstrate a comprehensive analysis of the transient NP array heating. Firstly, we developed analytical solution for the NP array heating and provide a useful mathematical description of the spatial-temporal evolution of temperature for 2D, 3D and spherical NP array heating. Based on this, we proposed the idea of thermal resolution that quantifies the relationship between minimal heating time, NP array size, energy intensity and target temperature. Lastly, we define a dimensionless parameter that characterize the transition from confined heating to delocalized heating. This study advances the in-depth understanding of nanomaterials heating and provides guidance for rationally designing innovative approaches for NP array heating.*

## KEY WORDS

*Nanoparticle array heating, thermal resolution, heating overlap, confined heating, delocalized heating*

## INTRODUCTION

Nanoparticles (NPs) have been applied as nanoscale heaters in a variety of applications[1-4], including thermal therapy[5-8], droplet heating[9], drug delivery[10-12], neuromodulation[13, 14], photoacoustic imaging[15, 16], photothermal imaging[17, 18], and photothermal catalysis[19-21]. For these applications, precise control of both the magnitude and spatiotemporal distribution of the temperature is often crucial[22, 23]. As an example, neuromodulation via the thermally sensitive ion channel TRPV1 requires a threshold temperature to activate the channel (40 °C – 53 °C)[24-26], while overheating (> 60 °C) can lead to cellular damage[27]. Additionally, for molecular hyperthermia, highly localized heating at the nanoscale is essential for targeted protein inactivation[28-30]. A comprehensive understanding of NP heating is particularly valuable when optimizing the spatiotemporal evolution of the temperature profile for a particular application. While single NP heating has been extensively investigated and is well understood[30-35], NP cluster or NP array heating is far from well understood due to the complexity of interactions among NPs and the diversity of possible NP array geometries.

Despite the inherent complexity of NP array heating, various works have shed light on the phenomena. For example, numerical and experimental investigations demonstrated that NP cluster heating may lead to a significantly larger temperature rise as compared to single NP heating, with the temperature rise for NP clusters increasing with NP concentration[36-41]. Additionally, efforts have been made to determine analytical solutions of NP array heating. Keblinski et al. derived a temperature function of 3D NP array heating under steady state[42]. They demonstrated that the array size and NP concentration determine the array temperature rise, which could be orders of magnitude higher than that of single NP heating due to heating overlap between NPs in the array. Baffou et al. further derived temperature functions for 1D and 2D NP array heating under steady state and repeated femtosecond pulsed excitations [43, 44]. Importantly, Baffou et al. described sets of dimensionless parameters that characterize the transition from confined heating to delocalized heating defined based the temperature functions[44]. Although these seminal works have improved our understanding of NP array heating under the steady state, a detailed description of transient NP array heating is lacking. Meanwhile, applications based on NP array heating under single pulsed heating with duration ranging from nanoseconds to minutes have been reported recently[24, 29, 45]. Consequently, an analysis of the spatiotemporal evolution of temperature during transient array heating is urgently needed.

In this report, we systematically analyzed the transient heating process for NP arrays. Firstly, we derived a set of analytical temperature functions that predict the temperature rise for three representative NP assemblies (2D, 3D, and spherical NP arrays (Figure 1)), alleviating the need to run costly numerical simulations. Next, we developed the concept of thermal resolution to quantify the minimal size and heating time required for a NP array to reach a specified target temperature rise at a given heating intensity, therefore providing guideline on physical limitations for applications based on NP array heating. Lastly, we defined a new set of dimensionless parameters based on interparticle distance and heating time to characterize the transition from confined heating to delocalized heating. From analysis based on these dimensionless parameters, we found that for spherical NP arrays, delocalized heating first occurs along the spherical surface, and then further delocalizes throughout the volume inside the sphere. This effect may be particularly relevant when heating NPs arrayed across the surface of a nanovesicle or living cell[10]. This work provides clear analytical guidance for designing innovative approaches that utilize NP heating under realistic physical constraints.

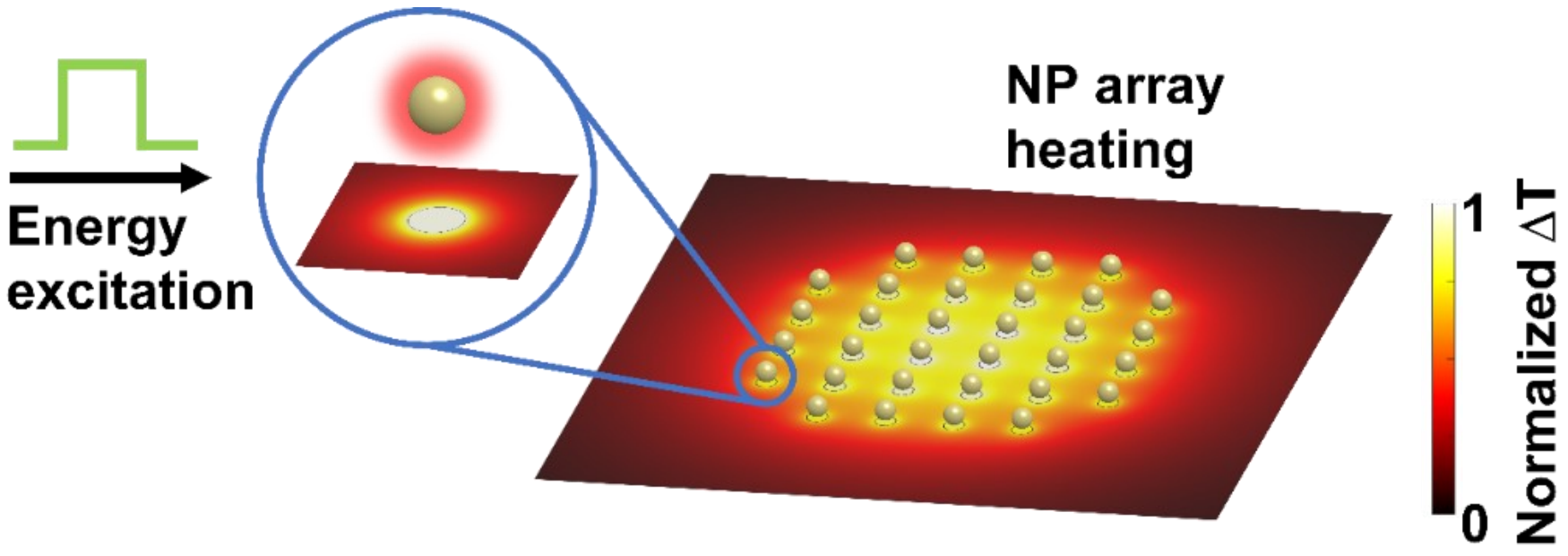

**Figure 1 Schematic illustrations of nanoparticle (NP) array.** Absorption of external energy excitation and conversion into heat by NP arrays.

## RESULTS

### Single NP heating

We start by considering single NP heating. Many of the applications of nanoparticle (NP) heating mentioned above are based on heating metallic NPs in aqueous solution. In these cases, NPs absorb external energy excitations, such as laser irradiation[46], magnetic[47] or ultrasonic fields[48], and convert that energy into heat (Figure 1)[1]. For a single NP, this nanoscopic absorption-heating process can be modeled by treating the NP as a spherical particle subject to volumetric heating and heat transfer into a surrounding homogeneous aqueous medium using an application of Fourier's law,

$$\begin{cases} \dfrac{1}{r^2}\dfrac{\partial}{\partial r}\left(r^2\dfrac{\partial T}{\partial r}\right)+\dfrac{q_v}{k_{NP}}=\dfrac{1}{\alpha_{NP}}\dfrac{\partial T}{\partial t},\ \ 0\le r<r_{NP},\ \ t\ge 0, \\ \dfrac{1}{r^2}\dfrac{\partial}{\partial r}\left(r^2\dfrac{\partial T}{\partial r}\right)=\dfrac{1}{\alpha}\dfrac{\partial T}{\partial t},\ \ r_{NP}\le r,\ \ t\ge 0, \\ T(r,t=0)=0,\ \ T(r=\infty,t)=T_\infty, \\ k_{NP}\dfrac{\partial T(r=r_{NP}^-,t)}{\partial r}=k\dfrac{\partial T(r=r_{NP}^+,t)}{\partial r} \end{cases} \tag{1}$$

where $T(\boldsymbol{r},t)$ is the temperature profile, $r$ is distance from NP center and $r_{NP}$ is radius of NP, $t$ is time, $q_V$ is volumetric heat source inside the NP, $\alpha_{NP}$ and $\alpha$, $k_{NP}$ and $k$ are thermal diffusivity and thermal conductivity for NP and water respectively. Although an analytical solution to Equation (1) can be derived by Laplace transform, known as Goldenberg's analytical solution[31], the complexity of the expression makes it challenging to use in our analysis of NP array heating. Therefore, we first simplify Equation (1) by assuming a constant heat flux at the NP-water interface:

$$\begin{cases} \dfrac{1}{r^2}\dfrac{\partial}{\partial r}\left(r^2\dfrac{\partial T}{\partial r}\right)=\dfrac{1}{\alpha}\dfrac{\partial T}{\partial t},\ \ r_{NP}\le r,\ \ \ t\ge 0, \\ T(r,t=0)=0,\ \ T(r=\infty,t)=T_\infty, \\ k\dfrac{\partial T(r=r_{NP},t)}{\partial r}+\dfrac{4}{3}\pi r_{NP}^3 q_v=0 \end{cases} \tag{2}$$

Equation (2) can be solved by Laplace transform[42]:

$$\Delta T_{single}=T-T_\infty=\frac{q}{4\pi rk}\left[erfc\left(\frac{r-r_{NP}}{2\sqrt{\alpha t}}\right)-exp\left(\frac{r-r_{NP}}{r_{NP}}+\frac{\alpha t}{r_{NP}^2}\right)erfc\left(\frac{r-r_{NP}}{2\sqrt{\alpha t}}+\frac{\sqrt{\alpha t}}{r_{NP}}\right)\right] \tag{3}$$

Note that by assuming a constant heat flux at the NP-water interface, we have neglected the heat capacity of the NP, and thus Equation (3) only describes the temperature profile in the surrounding water during single NP heating with applied power $q$ (heating power per NP, μW). Since the heat capacity of the NP is small relative to that of the surrounding water, this approximation should only lead to a minor overestimation of the water temperature at the start of heating[42], yet the expression in Equation (3) is much simpler than Goldenberg's analytical solution to Equation (1). Therefore, we will make use of Equation (3) in our analysis of NP array heating.

**NP array heating**

Most applications of NP heating are based on multi-NP heating or NP cluster heating with thermal interactions among individual NPs. The diverse set of cluster geometries that multi-NP systems can assume makes it challenging to derive analytical descriptions of multi-NP heating. To circumvent this issue, we narrowed our scope to three representative NP cluster geometries (2D, 3D and spherical NP cluster). And we modeled the multi NP heating via NP arrays with periodic lattices (Figure 2, square and cubic lattice for 2D and 3D array respectively, Fibonacci lattice for spherical array[49]) and assume uniform heating power throughout the array.

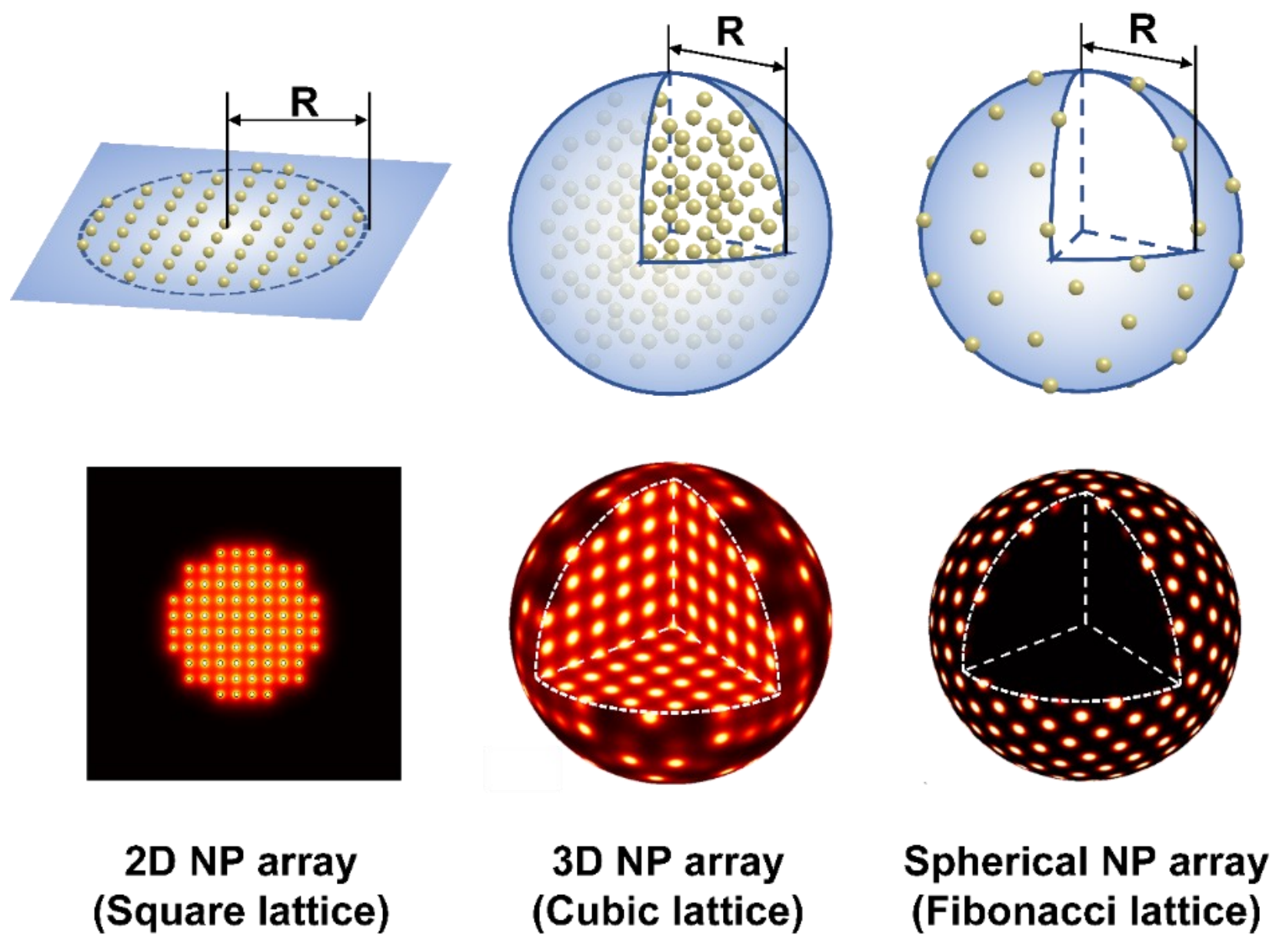

**Figure 2 Schematic illustrations of representative NP array geometries:** 2D NP array with square lattice, 3D NP array with cubic lattice, spherical NP array with Fibonacci lattice. For 2D and 3D NP array, $R$ represents the size of NP array; for spherical NP array, $R$ represents the size of the sphere.

The temperature rise in the NP array ($\Delta T_{array}$) can be estimated by superposition of temperature rise of individual NP heating ($\Delta T_{single}$)[44]:

$$\Delta T_{array}(\boldsymbol{r}, t) = \sum_{i=1}^{N} \Delta T_{single,i}(|\boldsymbol{r} - \boldsymbol{r}_i|, t) \tag{4}$$

where $N$ is the total number of NPs, $\boldsymbol{r}_i$ is the location for $i$th NP. For a circular shaped 2D NP array, the temperature rise at the center of the array ($\Delta T_{2D}$) can be estimated by:

$$\Delta T_{2D}(t) = \sum_{i=1}^{N} \Delta T_{single,i}(r_i, t) \tag{5}$$

where $r_i$ is the distance from the $i$th NP to the center of NP array. Figure 3A shows that, when the interparticle distance, $p$, is small relative to the size of the array, $R$, (i.e., $R/p \gg 1$), summation in Equation (5) can be estimated by an integration:

$$\Delta T_{2D}(R,t) = \int_{r=0}^{r=R} \Delta T_{single}(r,t) \cdot 2\pi r \rho_{NP} \cdot dr \tag{6}$$

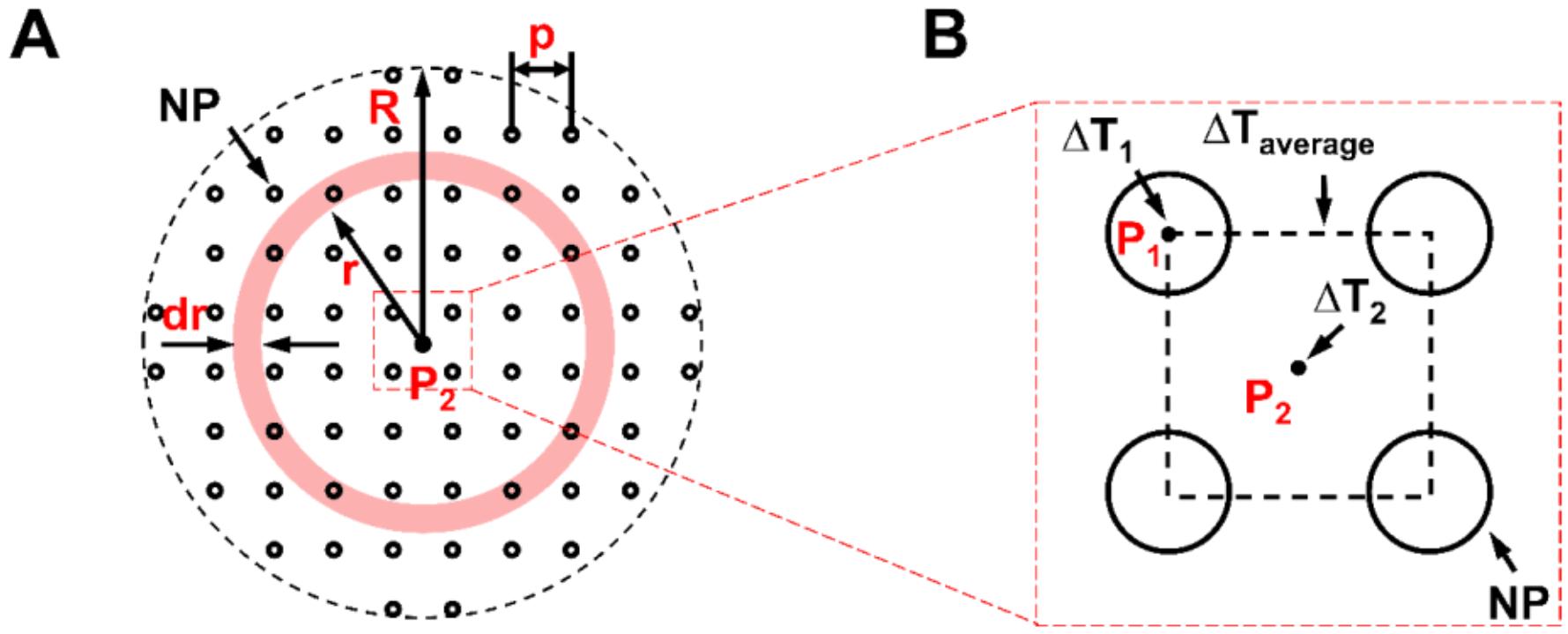

**Figure 3 Schematic illustrations of theoretical model:** (A) Temperature function for 2D NP array is derived by transforming super-position into integration. (B) Representative locations and temperatures in the center lattice of the NP array. $P_1$ is located in NP and $P_2$ is located in the mid-point between NPs.

Here $R$ is the size of the NP array (Figure 2&3A) and $\rho_{NP}$ is the NP concentration ($\mu m^{-2}$ for 2D and spherical NP array, $\mu m^{-3}$ for 3D NP array). Substitute Equation (3) into Equation (6) gives:

$$\Delta T_{2D} = \frac{q\rho_{NP}}{2k}\left[R \cdot erfc\left(\frac{R}{2\sqrt{\alpha t}}\right) + \frac{2\sqrt{\alpha t}}{\pi}\left(1 - exp\left(\frac{-R^2}{4\alpha t}\right)\right)\right] + r_{NP}\frac{q\rho_{NP}}{2k} \cdot \left\{exp\left(\frac{R}{r_{NP}} + \frac{\alpha t}{{r_{NP}}^2}\right) erfc\left(\frac{r}{2\sqrt{\alpha t}} + \frac{\sqrt{\alpha t}}{r_{NP}}\right) - \exp\left(\frac{\alpha t}{{r_{NP}}^2}\right) erfc\left(\frac{R}{2\sqrt{\alpha t}} + \frac{\sqrt{\alpha t}}{r_{NP}}\right) + erfc\left(\frac{R}{2\sqrt{\alpha t}}\right) - 1\right\} \tag{7}$$

Since the size of individual NP ($r_{NP}$) is minimal comparing with that of the NP array ($R \gg r_{NP}$), thus we ignore the second term of the righthand side of Equation (7):

$$\Delta T_{2D} = \frac{q\rho_{NP}}{2k}\left[R \cdot erfc\left(\frac{R}{2\sqrt{\alpha t}}\right) + 2\sqrt{\frac{\alpha t}{\pi}}\left(1 - exp\left(\frac{-R^2}{4\alpha t}\right)\right)\right] \quad (8)$$

Equation (8) describes temperature rise at the center point of a 2D NP array.

We validated Equation (8) by comparing it with simulation results (superposition method based on Goldengerg's analytical solution (Equation (S2))). It is notable that NP array heating with large interparticle distance and short heating time can result in heterogenous temperature profile (Figure S1), i.e. confined heating. Therefore, we compared Equation (8) to three representative $\Delta T$ values computed from numerical simulation of superposition (Figure 3B & S1): $\Delta T_1$, $\Delta T_2$, and $\Delta T_{average}$, which correspond to the temperature rise inside an NP (at P1), the mid-point between NPs (at P2), and the average throughout central lattice, respectively. Figure 4 shows that Equation (8) accurately describes temperature rise especially for $\Delta T_{average}$, even for small NP arrays ($R/p = 1$), and the accuracy of Equation (8) is confirmed.

Following similar process, we derived temperature functions for 3D NP array and spherical NP array (Table 1, Equations 9 and 10). Figure S2 shows that results from Equation (9) and Equation (10) are in good agreements with simulation results, which confirms the accuracy. Considering the identical distance from NPs on the spherical surface to the center point and the superposition method, Equation (11), which describes the temperature rise at the center point of a spherical NP array heating, is derived via multiplying Equation (3) by the total number of NPs ($N$).

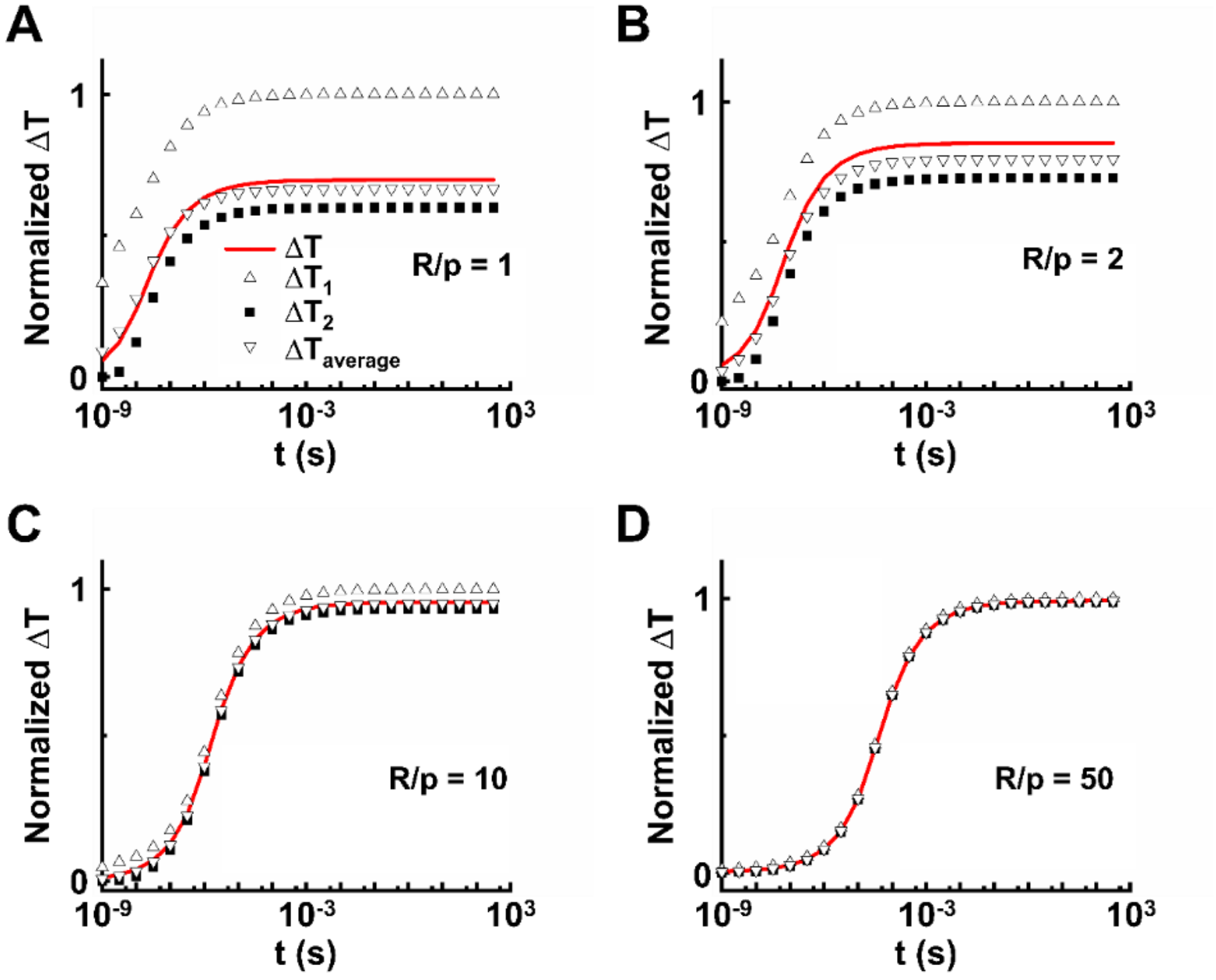

**Figure 4 Validation of the temperature function for 2D NP array heating:** Comparing the function with solution of super-position results. The results from superposition are calculated based on Goldenberg's analytical solution for single NP heating (Equation (S2)), $r_{NP}$ = 15 nm. The size of the NP array ($R$) comparing with interparticle distance ($p$) is (A) $R/p$ = 1, (B) $R/p$ = 2, (C) $R/p$ = 10, (D) $R/p$ = 50.

**Table 1 Temperature functions for 2D, 3D and spherical NP arrays.**

| NP array geometry | Location | Temperature function | |
|---|---|---|---|
| 2D NP array | Center point | $\Delta T_{2D} = \frac{q\rho_{NP}}{2k}\left[R \cdot erfc\left(\frac{R}{2\sqrt{\alpha t}}\right) + 2\sqrt{\frac{\alpha t}{\pi}}\left(1 - exp\left(\frac{-R^2}{4\alpha t}\right)\right)\right]$ | (8) |
| 3D NP array | Center point | $\Delta T_{3D} = \frac{q\rho_{NP}}{2k}\left[R^2 \cdot erfc\left(\frac{R}{2\sqrt{\alpha t}}\right) - 2\sqrt{\frac{\alpha t}{\pi}}R \cdot exp\left(\frac{-R^2}{4\alpha t}\right) + 2\alpha t \cdot erf\left(\frac{R}{2\sqrt{\alpha t}}\right)\right]$ | (9) |
| Spherical NP array | On spherical surface | $\Delta T_{SPH,s} = \frac{q\rho_{NP}}{k}\left[R \cdot erfc\left(\frac{R}{\sqrt{\alpha t}}\right) + \sqrt{\frac{\alpha t}{\pi}}\left(1 - exp\left(\frac{-R^2}{\alpha t}\right)\right)\right]$ | (10) |
| | Center point | $\Delta T_{SPH,c} = \frac{q\rho_{NP}}{k}\left[R \cdot erfc\left(\frac{R - r_{NP}}{\sqrt{\alpha t}}\right) - R \cdot exp\left(\frac{R - r_p}{r_{NP}} + \frac{\alpha t}{{r_{NP}}^2}\right) erfc\left(\frac{R - r_{NP}}{2\sqrt{\alpha t}} + \frac{\sqrt{\alpha t}}{r_{NP}}\right)\right]$ | (11) |

A close inspection shows that Equation (8)-(11) can be rewritten into similar mathematical form:

$$\Delta T = \frac{1}{2k} \cdot \underbrace{q\rho_{NP}}_{\text{Energy intensity term}} \cdot \underbrace{f(R,t)}_{\text{Spatiotemporal term}} \tag{12}$$

where $q\rho_{NP}$ serves as energy intensity term, and $f(R,t)$ is the spatiotemporal diffusion term. Equation (12) reveals the underlying principle for transient NP array heating: Temperature raise equals to a production of the linear energy intensity term, which is determined by heating power per NP and NP concentration, with a non-linear spatiotemporal term, which is dependent on the NP array geometry and heating time.

The simple analytical solutions (Equations 8-11) provide a method for further analysis on transient NP array heating as well as useful expressions to estimate temperature rise of representative NP array heating, alleviating the need to run heavy numerical simulations. For example, Equation (8) can be used to estimate temperature rise for lithographically fabricated NP heating with repeated 2D lattice[37]. Similarly, Equation (9) is useful in estimating temperature rise for photothermal therapy (PTT) or photothermal catalysis where free suspension of NPs in aqueous solution can be treated as 3D NP array[5, 9, 19].

**Application of NP array heating to reach a specified temperature change**

The spatiotemporal scale of the NP array heating is often crucial for its applications. As an example, for photothermal therapy (PTT) or plasmonic photothermal therapy (PPTT), a precise control of the heating range is necessary for effective tumor destruction while avoiding healthy tissue damage[7, 50]. On the other hand, specific temperature threshold is required to trigger a biological or chemical response. With these considerations, we analyzed the NP array heating from an alternative perspective: what are the necessary conditions for a specific temperature rise? Based on this, we analyzed the critical NP array size and critical heating time for a specific temperature rise, which quantifies the spatiotemporal scale for NP array heating.

We started with defining the target temperature rise ($\Delta T_{target}$) for a specific application, such as activating thermally sensitive ion channel or trigger tumor cell death[25, 27]. To evaluate the necessary conditions for $\Delta T_{target}$, we rewrite Equation (12) into:

$$f(R_c, t_c) = \Delta T_{target} \cdot \frac{2k}{q\rho_{NP}} \tag{13}$$

where $R_c$ and $t_c$ are critical NP array size and critical heating time. For a specific $\Delta T_{target}$ and energy intensity ($q\rho_{NP}$), Equation (13) describes a combination of critical NP array size ($R_c$) and heating time ($t_c$) to reach the specified temperature increase. In other words, for each heating time ($t_c$), a critical NP array size ($R_c$) can be determined (Figure 5) and is the minimal NP array size to reach $\Delta T_{target}$ under the specific heating time ($t_c$). Similarly, for a given NP array size ($R_c$), $t_c$ represents the minimal heating time to reach $\Delta T_{target}$. For example, Figure 5 shows that at least a 2D NP array with 1.3 μm in diameter is needed to reach 1 K temperature rise under 1 ms heating ($\rho_{NP}$ = 100 μm$^{-2}$, $q$ = 10 nW/NP).

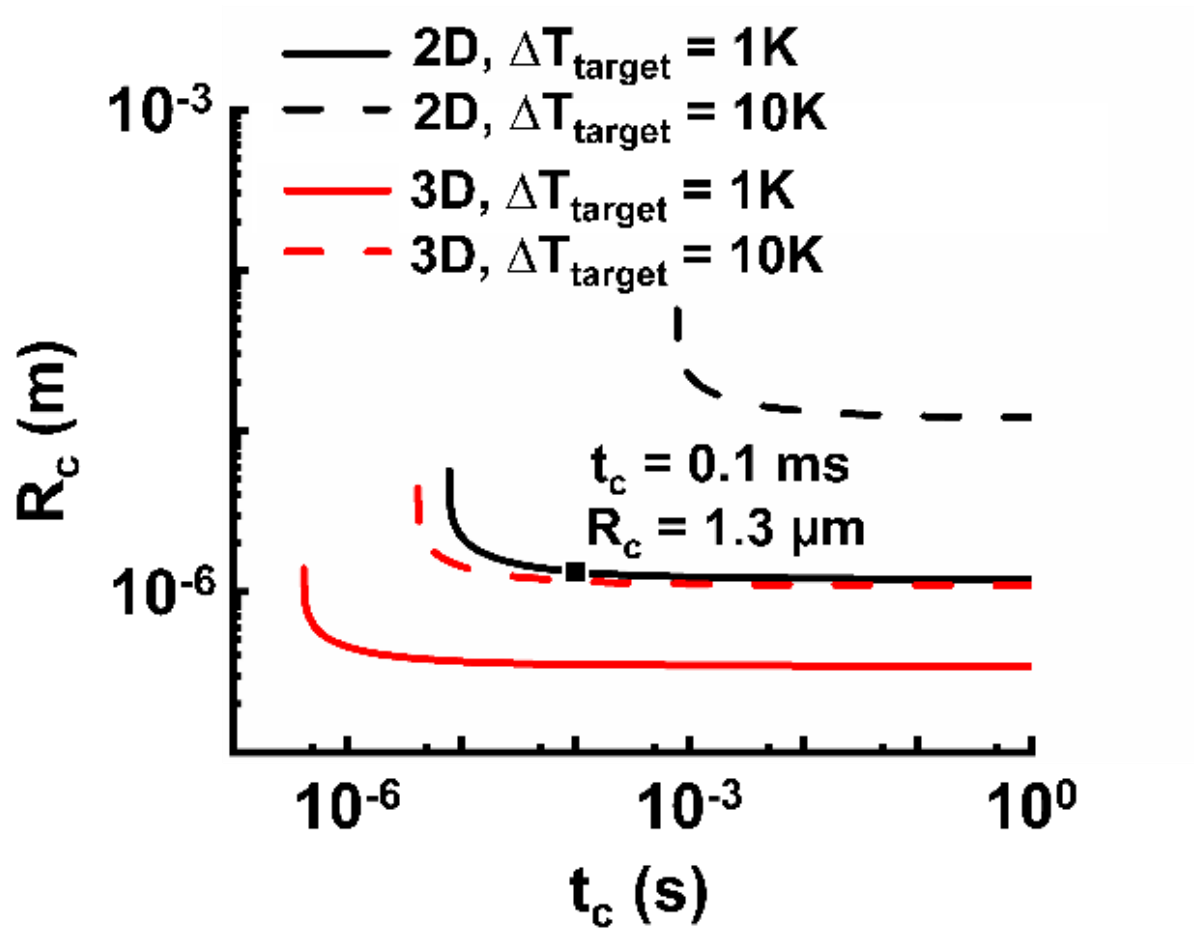

**Figure 5 Critical size of 2D and 3D NP arrays for specific temperature rise:** Critical NP array size $R_c$ and in terms of heating time $t_c$ for 2D and 3D NP array. NP concentration ($\rho_{NP}$): 100 $\mu m^{-2}$ for 2D array and 1000 $\mu m^{-3}$ for 3D array; heating power ($q$): 10 nW/NP.

Furthermore, the NP array size ($R_c$) reaches an asymptotic value when $t_c$ goes to infinity, a.k.a. under steady state (s.s.):

$$R_{c,s.s.} = \begin{cases} \dfrac{2k}{q\rho_{NP}} \cdot \Delta T_{target}, & 2D\ NP\ array \\ \sqrt{\dfrac{2k}{q\rho_{NP}} \cdot \Delta T_{target}}, & 3D\ NP\ array \end{cases} \quad (14)$$

Equation (14) and Figure 6A show the smallest array size to reach the required temperature change ($\Delta T_{target}$) under the steady state. Similarly, a larger sized NP array ($R_c$) can lead to a shorter heating time ($t_c$). With an infinitely large array, a minimum heating time can be obtained to reach the specified temperature change ($\Delta T_{target}$):

$$t_{c,inf.} = \begin{cases} \dfrac{\pi k^2}{\alpha}\left(\dfrac{1}{q\rho_{NP}} \cdot \Delta T_{target}\right)^2, & 2D\ NP\ array \\ \dfrac{k}{\alpha} \cdot \dfrac{1}{q\rho_{NP}} \cdot \Delta T_{target}, & 3D\ NP\ array \end{cases} \quad (15)$$

Equation (15) and Figure 6B show that minimal heating time with infinite sized NP array ($t_{c,inf.}$) is determined by energy intensity. For example, Figure 6B shows that, for 3D NP array heating with energy intensity at 1 pW/μm$^{-3}$, the heating time should be at least 1 s to reach a 1 K temperature rise.

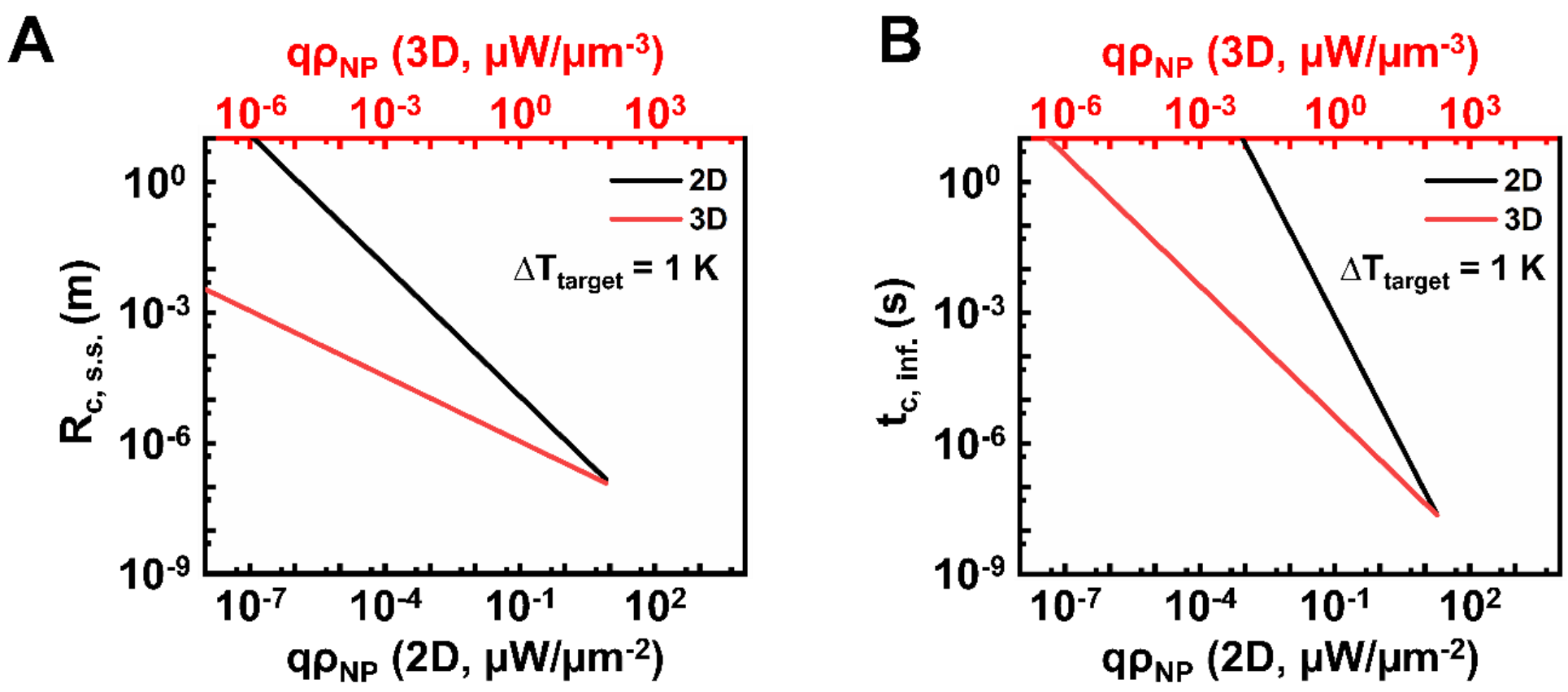

**Figure 6 Critical size under steady state (A) and critical heating time with infinite NPs (B) for 2D and 3D NP arrays in terms of energy intensity.** Unit for energy intensity is μW/μm$^{-2}$ for 2D NP arrays and μW/μm$^{-3}$ for 3D NP arrays.

## Thermal resolution

Our analysis of the critical NP array size ($R_c$) can be applied to define a "thermal resolution". An emerging application involves in activating neurons in the retina to restore vision[24]. Dasha et al. showed that, when excited by near-infrared radiation, the plasmonic gold nanorods attached on the retina serve as nano heaters, and can thermally activate TRPV1 channel. The heating of nanorods inside the light spot can be modeled as 2D NP array heating, where the size of NP array ($R$) is determined by the size of light spot projected on the retina. One of the key questions is that what is the resolution for this application from the heat transfer's perspective? As such, we define a "thermal resolution" based on the critical size of NP array heating ($R_c$). Here we assume a $R_c$ sized NP array heating is necessary to reach the temperature threshold for TRPV1 activation. Figure 7&8 shows that the minimal distance ($D = 2R_c$) at which two $R_c$ sized hot spots can be distinguished from each other, otherwise when $D < 2R_c$, the merged hot spots (Figure 8A) will lead to an overlap of TRPV1 activation, ending up with single "sensed" light spot. Consequently, $R_c$ serves as the spatial thermal resolution.

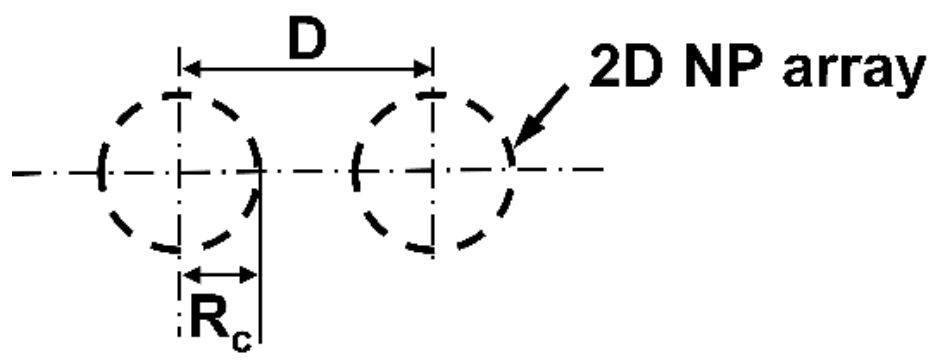

**Figure 7 Schematic illustration of inter array distance (*D*).**

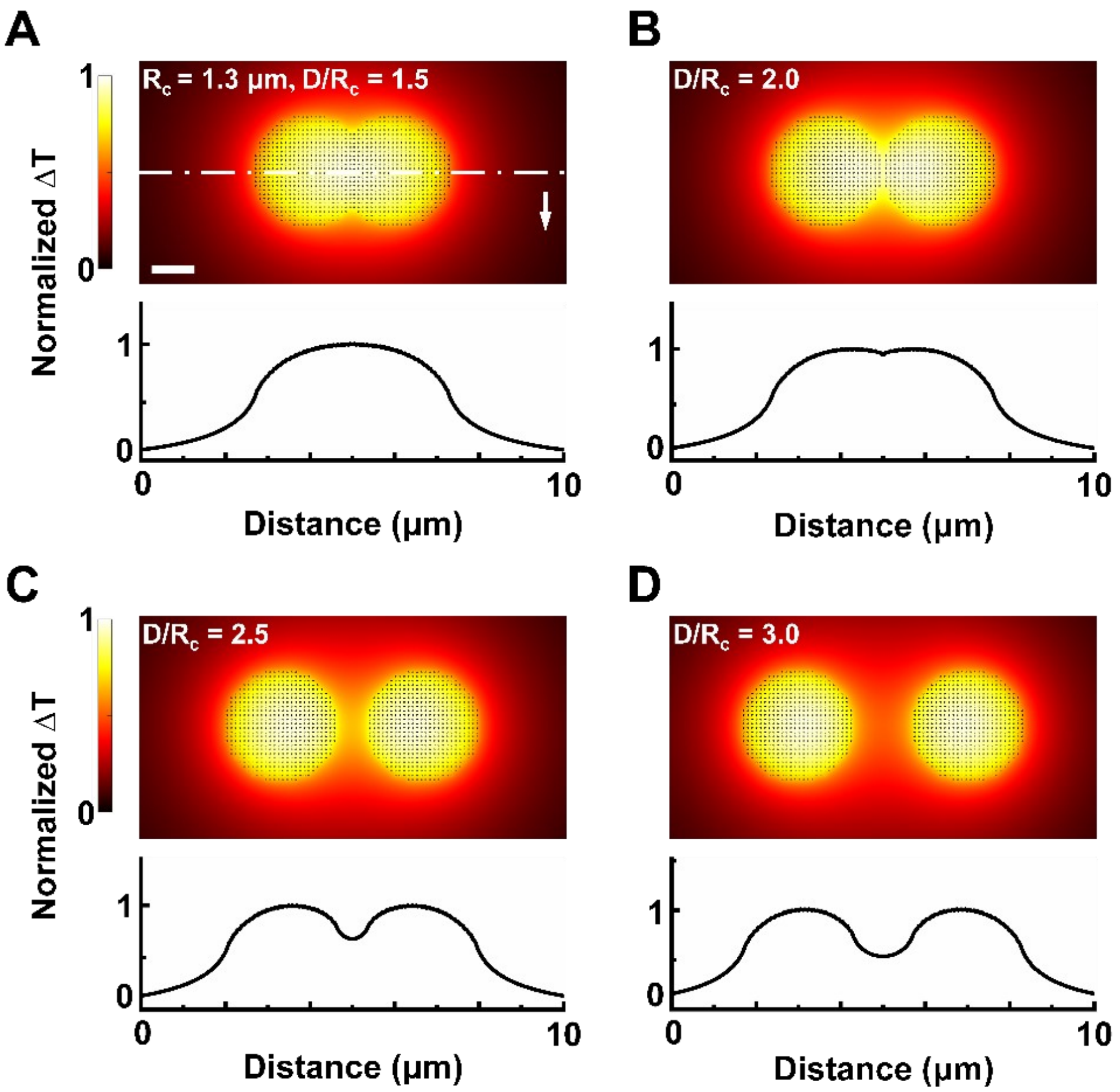

**Figure 8 Temperature profiles for two 2D NP arrays with different inter array distances (*D*):** $R_c$ = 1.3 µm, $t_c$ = 1 ms, $r_{NP}$ = 15 nm, $\rho_{NP}$ = 100 µm$^{-2}$, $q$ = 10 nW/NP, $D$ = (A) 1.95 µm (B) 2.6 µm (C) 3.25 µm (D) 3.9 µm. Scalebar represents 1 µm. When $D > 2R_c$, two distinguishable hot spots corresponding to the NP arrays are observed, while $D \leq 2R_c$, the two hot spots merged and can no longer be distinguished as individuals.

As mentioned above, the critical NP array size under steady state ($R_{c,s.s.}$) shows an upper limit for thermal resolution. Equation (14) and Figure 6A show that the $R_{c,s.s.}$ is determined by the energy intensity ($q\rho_{NP}$): higher energy intensity would result in greater possible thermal resolution (smaller $R_{c,s.s.}$). There are several significant implications from this result. Recently, various studies have focused on applications with localized NP heating, from targeted tumor thermal treatment[5], to neuron modulation[24] and molecular hyperthermia[29]. Our analysis shows that the thermal resolution gives a spatial limit when designing specific micro or nano heating patterns, and that the energy intensity is the key for greater thermal resolution. As an example, Figure 6A shows that an energy intensity ($q\rho_{NP}$) of 0.1 $\mu W/\mu m^{-3}$ is necessary for selective cellular heating ($R_c$ ~ 1-10 $\mu m$) with 3D NP array.

**Dimensionless parameter charactering confined heating**

In the previous section we discussed the degree of thermal confinement in the perspective of necessary condition for a NP array to reach the $\Delta T_{target}$ and developed the idea of thermal resolution. In this section, we will focus on thermal confinement inside the NP array i.e. heating overlap among NPs in a NP array.

One of the basic ideas for characterizing the degree of confined heating is to compare the thermal diffusion distance with the interparticle distance. Kang et al. developed dimensionless parameter based on comparing the thermal diffusion distance of transient single NP heating with the interparticle distance[30], which nicely quantifies the heating overlap between two NPs. Despite this, parameters that consider the ensemble effect among NPs in a NP array would greatly advance our understanding of the heating modes. Here we developed a set of dimensionless parameters to characterize confined heating based on quantifying the thermal diffusion distance in a NP array.

According to Equation (8)-(11), larger NP array size ($R$) gives greater $\Delta T(R,t)$, and when $R$ tends to infinity, a heating time dependent $\Delta T_{inf.}(t)$ can be expressed as following:

$$\Delta T_{inf.} = \begin{cases} \frac{q\rho_{NP}}{k}\sqrt{\frac{\alpha t}{\pi}}, & 2D\ NP\ array \\ \frac{q\rho_{NP}}{k}\alpha t, & 3D\ NP\ array \end{cases} \tag{16}$$

Figure 9A shows $\Delta T(R,t)$ in terms of $R$ for a specific heating time: Larger $R$ results in greater $\Delta T(R,t)$, which asymptotically reaches $\Delta T_{inf.}(t)$ as $R$ tends to infinity. Here a critical $R_{diff.}(t)$ is defined as when $\Delta T(R,t)$ approximates $\Delta T_{inf.}(t)$ (Figure 9A). As shown in Figure 9A, $R_{diff.}(t)$ divides the $R$ into R-limiting zone ($R < R_{diff.}(t)$), where $\Delta T(R,t)$ increases significantly with $R$, and the t-limiting zone ($R \geq R_{diff.}(t)$), where $\Delta T(R,t) \sim \Delta T_{inf.}(t)$ and no obvious increase for $\Delta T(R,t)$ is observed. The sharp contrast between these two zones demonstrates that $R_{diff.}(t)$ serves as the thermal diffusion distance in a NP array, where only NPs within the distance of $R_{diff.}(t)$ impact the temperature change at the center point (Figure 9B). It should be noticed that $R_{diff.}(t)$ is time dependent, where a increasing $t$ results in larger $R_{diff.}(t)$ (Figure S3).

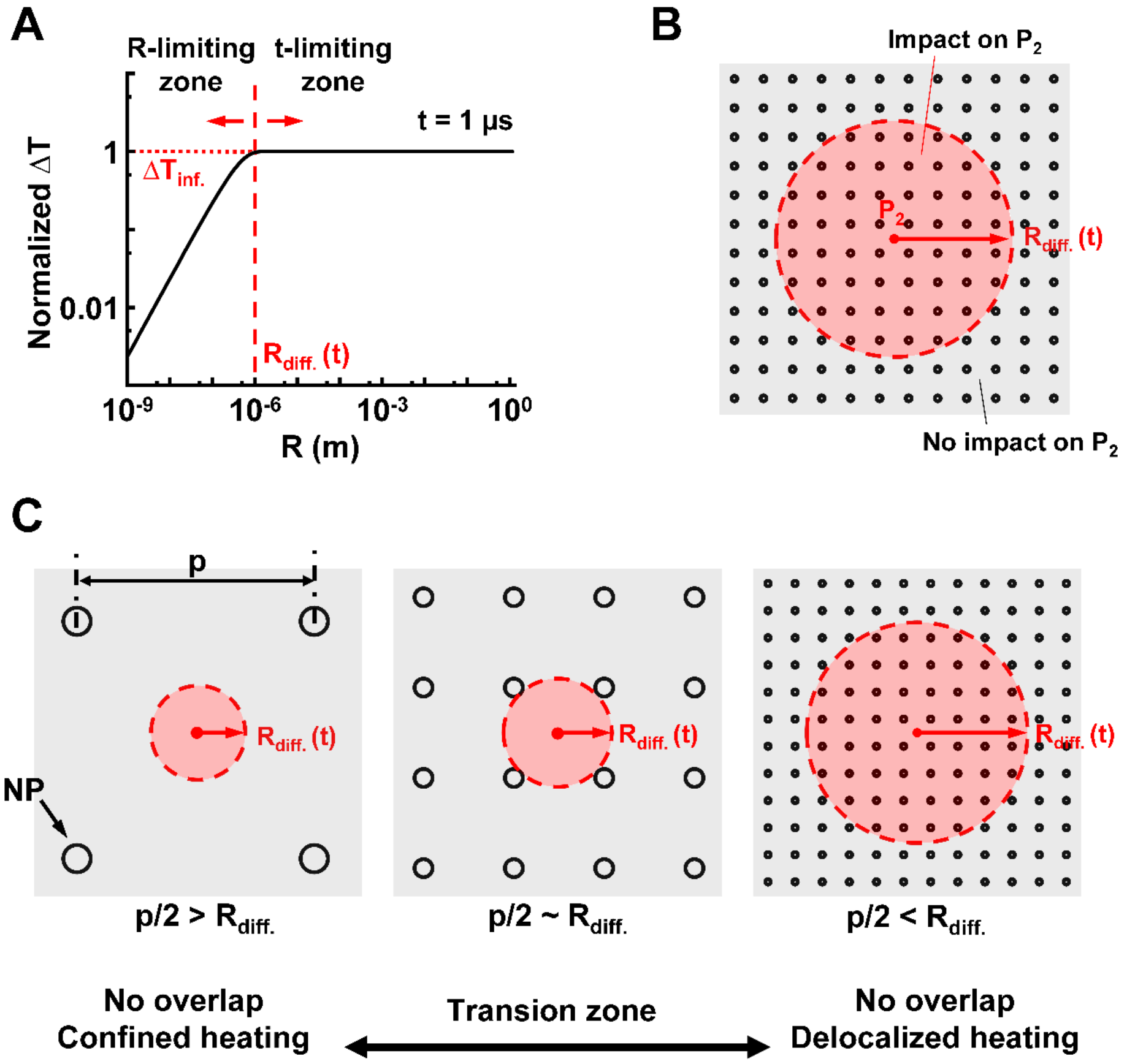

**Figure 9 Thermal diffusion distance in NP array ($R_{diff.}$).** (A) $\Delta T$ in terms of $R$ with a specific heating time (t = 1 μs). When $\Delta T$ approximates $\Delta T_{inf.}$, a critical NP array size can be observed ($R_{diff.}$). When $R < R_{diff.}$, increasing $R$ results in significant increase of $\Delta T$, while for $R > R_{diff.}$, $\Delta T$ is approximately equal to $\Delta T_{inf.}$. (B) $R_{diff.}$ serves as thermal diffusion distance where only the NPs within the distance of $R_{diff.}$ can thermally impact the center point. (C) We can characterize heating overlap by comparing $R_{diff.}$ with interparticle distance ($p$).

Given the thermal diffusion in a NP array, we can quantify the heating overlap and develop the parameter by comparing the $R_{diff.}(t)$ with interparticle distance ($p$) (Figure 9C): if $p/2 \gg R_{diff.}(t)$ ($2R_{diff.}/p \ll 1$), heating overlap is minimal and a confined heating is established; whereas if $p/2 \ll R_{diff.}(t)$ ($2R_{diff.}/p \gg 1$), significant heating overlap leads to delocalized heating. However, there are two key problems to develop a dimensionless parameter based on this idea:

(1) The mathematical definition of $R_{diff.}$ is not clear. As mentioned above, the $R_{diff.}$ is defined as when $\Delta T(R,t) \sim \Delta T_{inf.}(t)$, yet the approximation is not clear and slight change of the approximation (i.e. $\Delta T/\Delta T_{inf.} = 0.9$ or $\Delta T/\Delta T_{inf.} = 0.99$) can result in significant difference in $R_{diff.}$ as well as in $p/2R_{diff.}$. This makes it challenging to determine $R_{diff.}$ mathematically.

(2) No explicit expression for $R_{diff.}$ Since $R_{diff.}$ is defined by the behavior of $\Delta T(R,t)$, $R_{diff.}$ can only be expressed implicitly by Equation (13), whereas a parameter with explicit expression to characterize heating overlap is required.

To overcome these problems, we quantified the heating overlap from an alternative perspective: here we define a $R'_{diff.} = p/2$, and by substituting this $R'_{diff.}$ into Equation (8) or (9), a $\Delta T(R'_{diff.},t)$ (or $\Delta T(p/2,t)$) can be determined explicitly. And by comparing this $\Delta T(R'_{diff.},t)$ with $\Delta T_{inf.}(t)$, we can quantify the relation between $R_{diff.}$ and $p$, i.e. characterizing the heating overlap. Based on this idea, we define a set of dimensionless parameters ($\zeta(p,t)$) to characterize the heating overlap among NPs in 2D and 3D NP array (Table 2). Figure 10 shows that $\zeta$ nicely characterizes the degree of heating overlap among NPs and captures the transition from confined heating to delocalized heating: $\zeta \rightarrow 1$ indicates minimal heating overlap and $\zeta \rightarrow 0$ indicates heating overlap between NPs.

**Table 2 Dimensionless parameters charactering heating overlap.**

| NP array geometry | Description | Temperature function | |
|---|---|---|---|
| 2D NP array | Heating overlap between NPs | $$\zeta_{2D} = \frac{\Delta T\left(\frac{p}{2},t\right)}{\Delta T_{inf.}(t)} = \frac{\left(\frac{p}{2}\right)\cdot erfc\left(\frac{p}{4\sqrt{\alpha t}}\right) + 2\sqrt{\frac{\alpha t}{\pi}}\left(1 - exp\left(\frac{-p^2}{16\alpha t}\right)\right)}{2\sqrt{\frac{\alpha t}{\pi}}}$$ | (17) |
| 3D NP array | | $$\zeta_{3D} = \frac{\Delta T\left(\frac{p}{2},t\right)}{\Delta T_{inf.}(t)} = \frac{\left(\frac{p}{2}\right)^2\cdot erfc\left(\frac{p}{4\sqrt{\alpha t}}\right) - \sqrt{\frac{\alpha t}{\pi}}p\cdot exp\left(\frac{-p^2}{16\alpha t}\right) + 2\alpha t\cdot erf\left(\frac{p}{4\sqrt{\alpha t}}\right)}{2\alpha t}$$ | (18) |
| Spherical NP array | Heating overlap between NPs along spherical surface | $$\zeta_{SPH,s} = \frac{\Delta T\left(\frac{p}{2},t\right)}{\Delta T_{inf.}(t)} = \frac{1}{2}p\sqrt{\frac{\pi}{\alpha t}}\cdot erfc\left(\frac{p}{2\sqrt{\alpha t}}\right) + 1 - exp\left(\frac{-p^2}{4\alpha t}\right)$$ $$\rho_{NP} = \frac{N}{4\pi R^2},\ p = \sqrt{\frac{1}{\rho_{NP}}}$$ | (19) |
| | Heating overlap between spherical surface and center point | $$\eta_{SPH,c} = 1 - \frac{\Delta T_{SPH,c}}{\Delta T_{SPH,s}} = 1 - \frac{R\cdot\left[erfc\left(\frac{R - r_{NP}}{\sqrt{\alpha t}}\right) - exp\left(\frac{R - r_{NP}}{r_{NP}} + \frac{\alpha t}{{r_{NP}}^2}\right)erfc\left(\frac{R - r_{NP}}{2\sqrt{\alpha t}} + \frac{\sqrt{\alpha t}}{r_{NP}}\right)\right]}{R\cdot erfc\left(\frac{R}{\sqrt{\alpha t}}\right) + \sqrt{\frac{\alpha t}{\pi}}\left[1 - exp\left(\frac{-R^2}{\alpha t}\right)\right]}$$ | (20) |

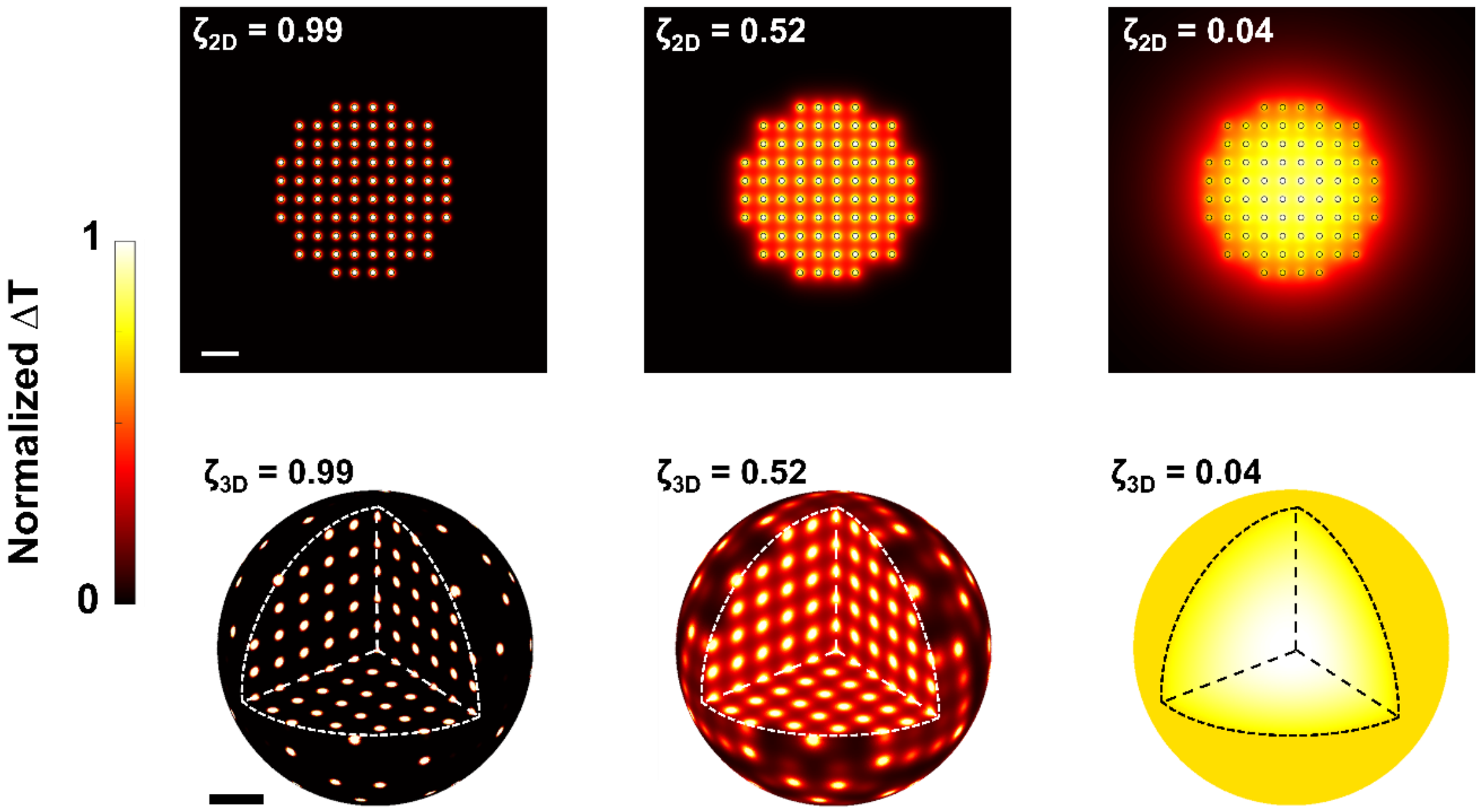

**Figure 10 Dimensionless parameter (ζ) characterizes the heating overlap and the transition from confined heating to delocalized heating in 2D and 3D NP array.** When $\zeta \sim 1$, confined heating is observed, whereas $\zeta \sim 0$ heating overlap leads to delocalized heating. Scalebar represents 200 nm, $r_{NP}$ = 15 nm.

For spherical NP array, parameter characterizing heating overlap along the spherical surface ($\zeta_{SPH,s}(p,t)$) is defined following a similar mathematical process (Table 2, Equation (19)); whereas parameter characterizing heating overlap between spherical surface and center point ($\eta_{SPH,c}(R,t)$) is defined by comparing the temperature rise on the spherical surface with that at the center point (Table 2, Equation (20)). A further inspection shows that the heating time ($t$) corresponding to $\zeta_{SPH,s}$ = 0.5 is constantly shorter than that corresponding to $\eta_{SPH,c}$ = 0.5 (Figure 11A), indicating heating overlap first occurs along the spherical surface, and then further delocalizes throughout the volume inside the sphere (Figure 11B), i.e., there is a window where heating is confined to the spherical surface with minimal heating inside the sphere; this effect may be particularly relevant when heating NPs arrayed across the surface of a nanovesicle or living cell.

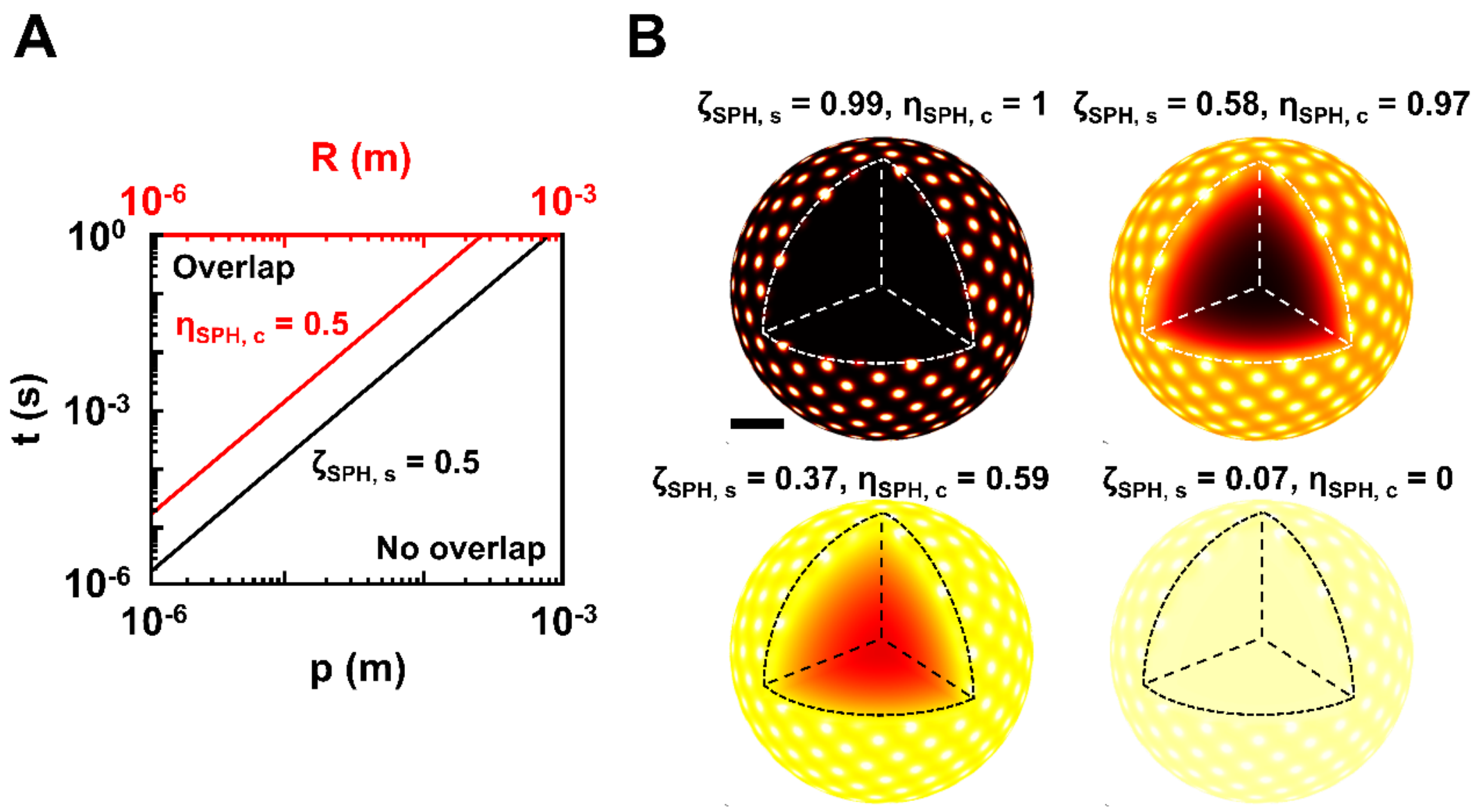

**Figure 11 Dimensionless parameters ($\zeta$ and $\eta$) characterize heating overlap and the transition from confined heating to delocalized heating for spherical NP array:** (A) Critical line for $\zeta_{SPH,s}$ = 0.5 and $\eta_{SPH,c}$ = 0.5. The critical line for $\eta_{SPH,c}$ = 0.5 is constantly above the line for $\zeta_{SPH,s}$ = 0.5, indicating delocalized heating first occurs along the spherical surface. (B) $\Delta T$ profile for spherical arrays. $\zeta_{SPH,s}$ and $\eta_{SPH,c}$ characterize the degree of heating overlap alone the spherical surface and inside the sphere. Scalebar represents 200 nm, $r_{NP}$ = 15 nm.

The dimensionless parameters we have derived are a significant step forward in differentiating confined heating from delocalized heating in the case of transient NP array heating, as they allow quantifying the effect of thermal spatiotemporal diffusion on heat confinement for a specific NP array. Recently, great progress has been made in developing novel applications based on nanoscale localized NP heating, such as molecular hyperthermia and neuron modulation[24, 29]. We anticipate that our work here advances the fundamental understanding of NP heating that underlies such applications, and provides a much-needed description of NP array heating that can be used when optimizing the spatiotemporal evolution of the temperature profile in these types of applications.

## DISCUSSION

### Physical meaning of the dimensionless parameter ($\zeta$)

The physical meaning of the $\zeta$ requires some further discussions. Based solely on the mathematical expressions, $\zeta$ represents the ratio of the temperature rise of a finite sized NP array to the temperature rise of an infinite sized NP array. However, neither of these descriptions truly represents the array we are actually analyzing. If we consider how $\zeta$ was derived, that suggests that it can be thought of as a transformed ratio between the interparticle distance ($p$) and the thermal diffusion distance ($R_{diff.}$). In other words, instead of directly comparing the two distances ($p$ and $R_{diff.}$), we substituted the distance into the spatiotemporal term (Equation (12)) and compared the results. In this case, the result from Equation (12) should be considered a transformed distance based on the spatiotemporal property of the NP array rather than the temperature rise. This interpretation seems reasonable for $\zeta_{2D}$ and $\zeta_{3D}$. However, this interpretation does not apply to $\zeta_{SPH,s}$ because the $R_{diff.}$ for a spherical array no longer represents a thermal diffusion distance when considering heat transfer along the spherical surface of the array.

**Relations between thermal resolution (*$R_c$*) and dimensionless parameter (*$\zeta$*)**

Both thermal resolution $R_c$ and dimensionless parameter $\zeta$ focus on the spatiotemporal distribution of the temperature profile. The thermal resolution is derived from an externally imposed target value of the temperature rise, which in turn leads to corresponding limits on the spatial and temporal scale of the NP array heating. In contrast, the dimensionless parameter reflects the internal shape of the temperature profile within a NP array, regardless of the magnitude of the temperature.

## SUMMARY

In this work, we aimed at a comprehensive analysis of the transient heating process for NP arrays by: (i) Deriving analytical temperature functions to predict the temperature rise at critical points for 2D, 3D, and spherical NP arrays, (ii) analyzing the conditions necessary to yield a specific temperature rise and developing the corresponding idea of thermal resolution, (iii) developing a set of dimensionless parameters that characterize the transition from confined heating (minimal heating overlap) to delocalized heating (significant heating overlap). In the case of the spherical NP array, our analysis revealed a window where there is substantial heating along the spherical surface but minimal heating inside the sphere, which we believe may be particularly relevant when designing applications based on heating NPs arrayed across the surface of nanovesicles or living cells. This work provides an in-depth understanding of the spatiotemporal evolution of temperature for NP array heating and provides analytical guidance for designing approaches based on this form of NP heating.

**ACKNOWLEDGMENT**

I am grateful for useful suggestions from Dr. Kang, Peiyuan (UTDallas), Dr. Wilson, Blake (UTDallas), Dr. Ye, Haihang (UTDallas) and Chaoran Dai (XJTU).

**FUNDING**

The research reported in this work was partially supported by the National Institute of General Medical Sciences (NIGMS) of the National Institutes of Health (award number R35GM133653), the Collaborative Sciences Award from the American Heart Association (award number 19CSLOI34770004), and the High-Impact/High-Risk Research Award from the Cancer Prevention and Research Institute of Texas (award number RP180846). The content is the sole responsibility of the authors and does not necessarily represent the official views of the funding agencies.

## NOMENCLATURE

### Roman letters

| | |
|---|---|
| $D$ | Inter array distance, m |
| $k$ | Thermal conductivity of water, W m$^{-2}$ K$^{-1}$ |
| $k_{NP}$ | Thermal conductivity of nanoparticle, W m$^{-2}$ K$^{-1}$ |
| $N$ | Total number of nanoparticles in an array |
| $p$ | Interparticle distance, m |
| $q$ | Heating power per nanoparticle, W |
| $q_v$ | Volumetric heating source, W m$^{-3}$ |
| $R$ | Radius of NP array, m |
| $r$ | Distance from the center, m |
| $\boldsymbol{r}$ | Location vector |
| $R_c$ | Spatial thermal resolution, M |
| $R_{c,s.s.}$ | Spatial thermal resolution under steady state, M |
| $R_{diff.}$ | Thermal diffusion distance, M |
| $R'_{diff.}$ | Substitute thermal diffusion distance, M |
| $r_{NP}$ | Radius of nanoparticle, m |
| $T$ | Temperature, K |
| $T_\infty$ | Room temperature, K |
| $\Delta T$ | Temperature rise, K |
| $\Delta T_1$ | Temperature rise in nanoparticle of the center lattice, K |
| $\Delta T_2$ | Temperature rise at the mid-point between nanoparticles of the center lattice, K |
| $\Delta T_{average}$ | Average temperature throughout the center lattice in a nanoparticle array, K |
| $\Delta T_{single}$ | Temperature rise of single nanoparticle heating, K |
| $\Delta T_{2D}$ | Temperature rise of 2D nanoparticle array heating, K |
| $\Delta T_{3D}$ | Temperature rise of 3D nanoparticle array heating, K |
| $\Delta T_{SPH}$ | Temperature rise of spherical nanoparticle array heating, K |
| $\Delta T_{target}$ | Specific temperature rise threshold for an application, K |
| $\Delta T_{inf.}$ | Temperature rise with infinite sized nanoparticle array heaing K |
| $t$ | Time, s |
| $t_c$ | Critical heating time, s |
| $t_{c,inf.}$ | Critical heating time with infinite sized nanoparticle array heaing, s |

**Greek symbols**

| | |
|---|---|
| $\alpha$ | Thermal diffusivity of water, $m^2\ s^{-1}$ |
| $\alpha_{NP}$ | Thermal diffusivity of nanoparticle, $m^2\ s^{-1}$ |
| $\rho_{NP}$ | Nanoparticle concentration, $m^{-2}$ (for 2D and spherical array), $m^{-3}$ (for 3D array) |
| $\eta_{SPH,c}$ | Dimensionless parameter for spherical nanoparticle array, at the center of the sphere |
| $\zeta_{2D}$ | Dimensionless parameter for 2D nanoparticle array |
| $\zeta_{3D}$ | Dimensionless parameter for 3D nanoparticle array |
| $\zeta_{SPH,s}$ | Dimensionless parameter for spherical nanoparticle array, on spherical surface |

**Abbreviations**

| | |
|---|---|
| SPH | Spherical |
| *NP* | Nanoparticle |
| *s.s.* | Steady state |
| inf. | Infinity |
| diff. | diffusion |
| TRPV1 | Transient receptor potential cation channel, subfamily V, member 1 |
| PTT | Photothermal therapy |
| PPTT | Plasmonic photothermal therapy |

## Supporting information for

# The Spatiotemporal Evolution of Temperature During Transient Heating of Nanoparticle Arrays

**This PDF file includes:**

Equations S1 to S2
Figures S1 to S3

## ANALYTICAL SOLUTION FOR SINGLE PARTICLE HEATING

The governing equation of single spherical particle heating is:

$$\begin{cases} \dfrac{1}{r^2}\dfrac{\partial}{\partial r}\left(r^2\dfrac{\partial T}{\partial r}\right)+\dfrac{q_v}{k_{NP}}=\dfrac{1}{\alpha_{NP}}\dfrac{\partial T}{\partial t}, \quad 0 \le r < r_{NP}, \quad t \ge 0, \\ \dfrac{1}{r^2}\dfrac{\partial}{\partial r}\left(r^2\dfrac{\partial T}{\partial r}\right)=\dfrac{1}{\alpha}\dfrac{\partial T}{\partial t}, \quad r_{NP} \le r, \quad t \ge 0, \\ T(r,t=0)=0, \quad T(r=\infty,t)=T_\infty, \\ k_{NP}\dfrac{\partial T(r=r_{NP}^-,t)}{\partial r}=k\dfrac{\partial T(r=r_{NP}^+,t)}{\partial r} \end{cases} \tag{S1}$$

And the analytical solution for Equation (1) is given be Laplace transform [1]:

$$\Delta T(r,t) = \begin{cases} \dfrac{r_{NP}{}^2 q_v}{k_{NP}}\left[\dfrac{1}{3}\dfrac{k_{NP}}{k} + \dfrac{1}{6}\left(1 - \dfrac{r^2}{r_{NP}{}^2}\right) - \dfrac{2r_{NP}b}{r\pi}\displaystyle\int_0^\infty \dfrac{\exp\left(-y^2 t/\gamma_1\right)}{y^2}\dfrac{(\sin y - y\cos y)\sin\left(ry/r_{NP}\right)}{[(c\sin y - y\cos y)^2 + b^2 y^2 sin^2 y]}dy\right], & r < r_{NP} \\ \dfrac{r_{NP}{}^3 q_v}{r k_{NP}}\left[\dfrac{1}{3}\dfrac{k_{NP}}{k} - \dfrac{2}{\pi}\displaystyle\int_0^\infty \dfrac{\exp\left(-y^2 t/\gamma_1\right)}{y^3}\dfrac{(\sin y - y\cos y)[by\sin y\cos\sigma y - (c\sin y - y\cos y)\sin\sigma y]}{[(c\sin y - y\cos y)^2 + b^2 y^2 sin^2 y]}dy\right], & r \geq r_{NP} \end{cases} \quad \text{(S2)}$$

$$b = \frac{k_{water}}{k_{gold}}\sqrt{\frac{\alpha_{Au}}{\alpha_{water}}}, c = 1 - \frac{k_{water}}{k_{gold}}, \gamma_1 = \frac{R_{NP}{}^2}{\alpha_{gold}}, \sigma = \left(\frac{r}{R_{NP}} - 1\right)\sqrt{\frac{\alpha_{gold}}{\alpha_{water}}}$$

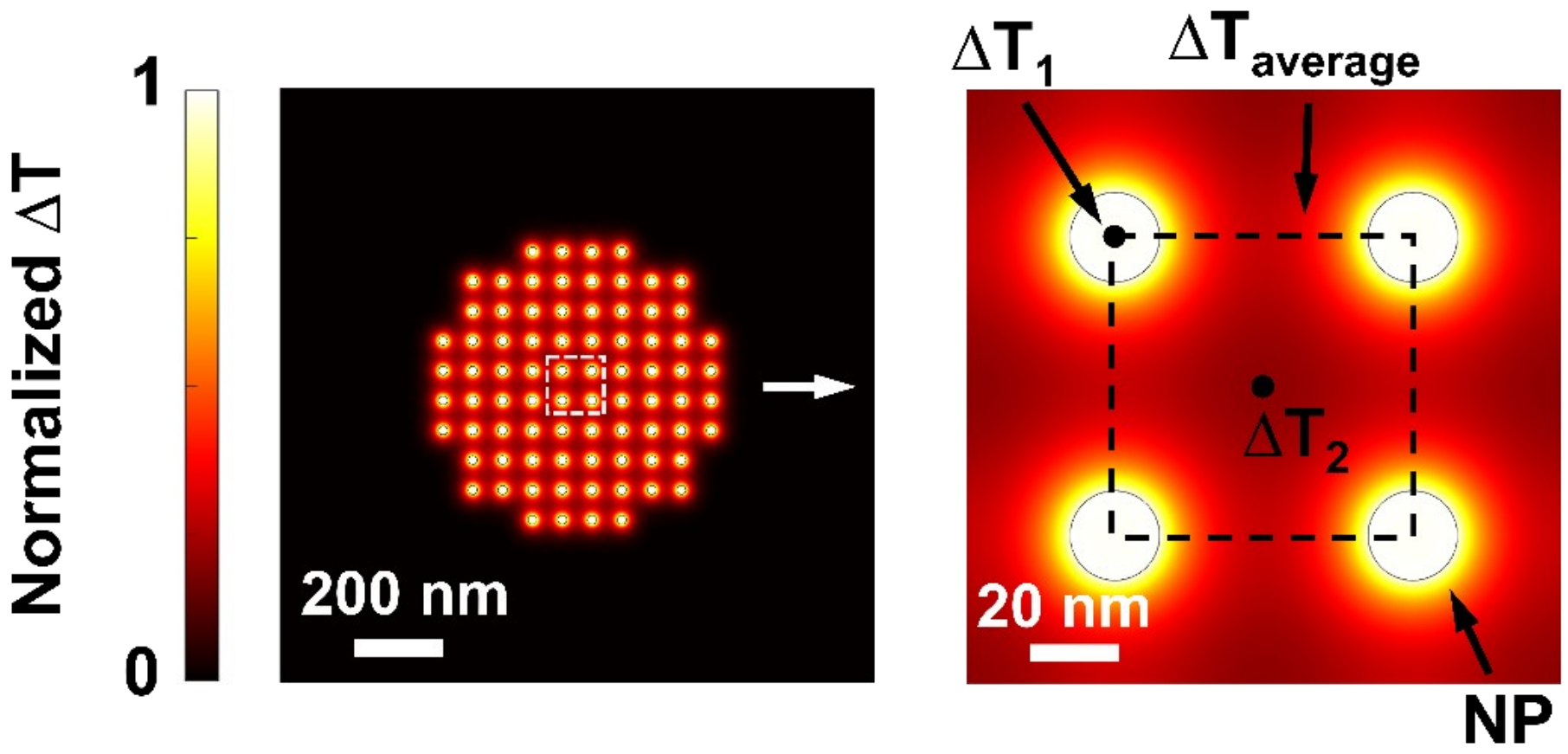

**Figure S1**. **Heterogeneous temperature profile for 2D NP array with short heating time:** The result from superposition is calculated based on Goldenberg's analytical solution for single NP heating (Equation (S2)), $p$ = 100 nm, $t$ = 10 ns, $r_{NP}$ = 15 nm.

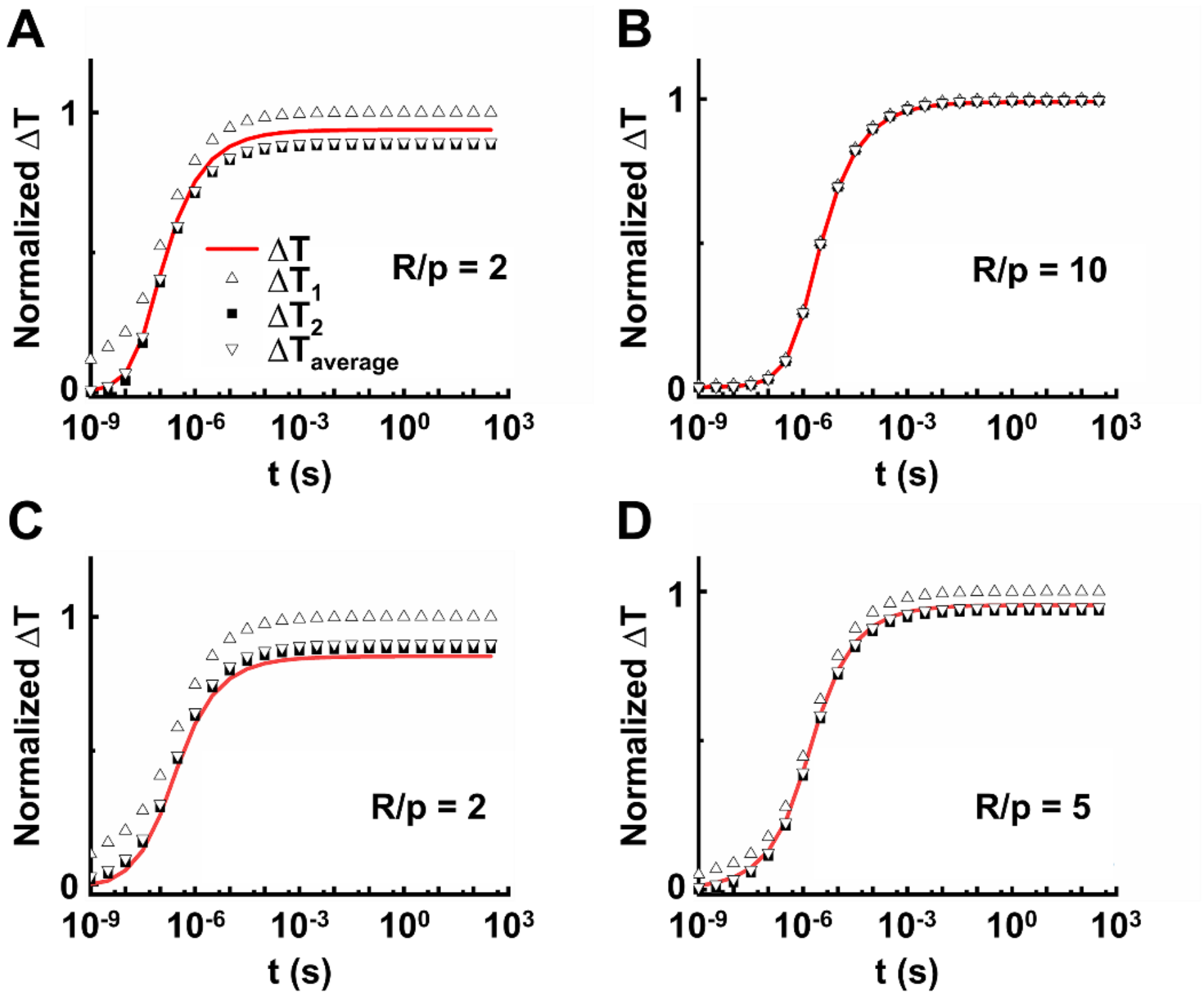

**Figure S2**. **Validation of the temperature function for 3D&SPH NP array heating:** (A&B) Validation of temperature function for 3D NP array (Equation (9)) by comparing the solution with super-position results. (C&D) Validation of temperature function for SPH NP array (Equation (10)) by comparing the solution with super-position results. The results from superposition is calculated based on Goldenberg's analytical solution for single NP heating (Equation (S2)), $r_{NP}$ = 15 nm.

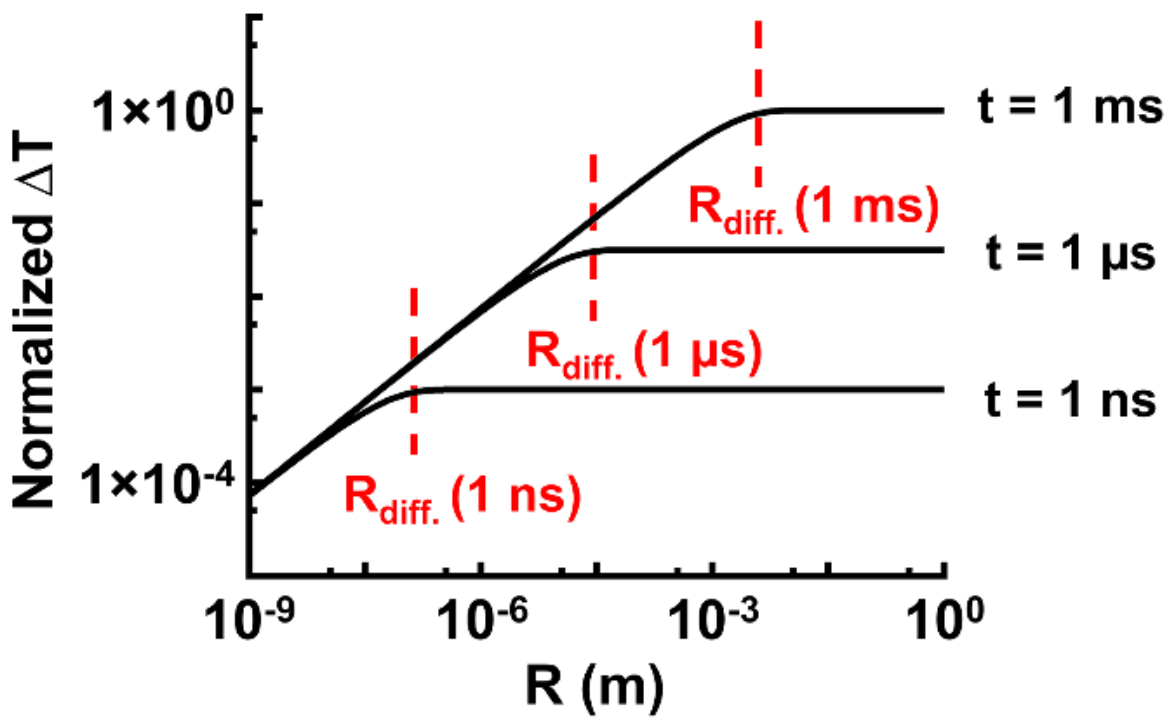

**Figure S3**. **Thermal diffusion distance in NP array ($R_{diff.}$) with different heating time ($t$).**